\documentstyle[twocolumn,prl,aps,epsf]{revtex}

\begin{document}        
\draft

\title {A Field Theory for Finite Dimensional \\
 Site Disordered Spin Systems }

\author{David S. Dean and David Lancaster}
\address{Dipartimento di Fisica and INFN, 
Universit\`a di Roma {\em La Sapienza}\\
P. A. Moro 2, 00185 Roma (Italy)}

\date{\today}

\maketitle

 
\begin{abstract}

We present a new field theoretic approach for finite dimensional site 
disordered spin systems by introducing  the notion of grand canonical 
disorder, where the number of spins in the system is random but quenched.
We perform the simplest non-trivial analysis of this field theory 
by using the variational replica formalism. 
We explicitly discuss a three dimensional RKKY-like system 
where we find a spin glass phase
with continuous replica symmetry breaking.

\end{abstract}
 

\narrowtext

Most advances in the field of disordered spin systems
have been based on models in which the bonds take random values.
However in most experimental systems the positions of the spins are random
but the interactions occur through deterministic potentials.
Analytic studies of  site-disordered spin systems,
such as RKKY spin glasses and dilute ferromagnets, 
have been hampered by the lack of a suitable field theoretic model (however
a lattice based formulation has been proposed \cite{Nie}).
By considering a situation in which 
the number of spins in the system is random but quenched
we are able to write a replica field theory for site-disordered systems.
This field theory seems to be simpler than many of those coming from 
bond-disordered and diluted lattice models and should be accessible
to many standard analytical techniques.
The mean field theory of this model, for reasons that will become clear, 
cannot provide any information about spin glass order.
In the second part of this letter we consider 
the simplest generalisation of mean field
theory, the Gaussian variational (GV) method, which does
provide this information.
Use of the GV method is widespread, and it is a useful warning
that for certain interaction types in our model it 
gives unphysical predictions.

The role of replica symmetry breaking (RSB) 
in disordered spin systems is of great interest.
Although RSB in the mean field 
theory for spin glasses is now well understood \cite{MePaVi} and related
to the proliferation of pure states of the system, 
in finite dimensions the picture is less clear.
Alternative qualitative approaches based on droplets\cite{FiHu} view the
spin glass phase as a disguised ferromagnetic phase with only 
two underlying fundamental states.
We will explicitly consider a 3-dimensional site disordered
spin glass using an RKKY-like interaction in our model and
find continuous RSB in the GV approximation.

Although in this letter we concentrate our attention on 
spin glass physics with oscillating sign interactions,
the model can also describe dilute
ferromagnetic or antiferromagnetic systems. Indeed, even for the
RKKY example we find ferromagnetic order at very low temperatures.
The application of the methods described here to these single
sign interactions is an interesting subject, but we defer it
to a longer article \cite{DeLa1}, simply mentioning some
of the issues that arise at the end of this letter.


Firstly consider a model where the number of spins $N$ is fixed: 
$N$ spins $S_i$ are placed randomly at positions $r_i$ 
uniformly throughout a volume $V$.
This type of disorder we  refer to as 
canonical disorder, as the number of particles is the same for each
realization of the disorder. The spins interact via a pairwise potential $J$
depending only on the distance 
between the spins. The Hamiltonian is then given by 
\begin{equation} H = -{1\over 2}\sum_{i,j} J(r_i - r_j)S_i S_j .
\label{Hamiltonian}\end{equation}
Assuming that $J$ is positive definite, making
a Hubbard-Stratonovich transformation expresses the partition function as 
\begin{eqnarray}
 Z_N &=& 
\sum_{S_i}\int \!{\cal D} \phi 
[det J\beta]^{-{1\over 2}} \times \label{ZedN} \\
&{\rm exp}& \left( {-1\over 2 \beta}\int\!\!\int  
\phi(r)J^{-1}(r-r')\phi(r')\ drdr'
+\sum_i^N\phi(r_i)S_i \right)\nonumber
\end{eqnarray}
Employing replicas,  
we average out the site-disorder by integrating over the
positions  $r_i$ using the flat measure: 
${1\over V^N}\int_V \prod dr_i $. 
\begin{eqnarray}
 {\overline Z_N^n}  & =&
\int \!{\cal D} \phi_a 
[det J\beta]^{-{n\over 2}}\times \nonumber \\
&{\exp}& \biggr( -{1\over 2 \beta}\int\!\int \sum_a 
\phi_a(r)J^{-1}(r-r')\phi_a(r')\ drdr' \nonumber\\
& + & N\log {1\over V}\int \prod^n_a 2\cosh \phi_a(r)\ dr \biggl)
\label{eq:can}
\end{eqnarray}
A field theoretic analysis of the above theory is complicated by the
presence of the $\log$ term in the action. 
We overcome this difficulty by making
a physically desirable modification to the definition of the disorder. In 
general one might expect the system to have been taken 
from a much larger system with a mean concentration of spins 
per unit volume, $\rho$. 
A suitably large  
subsystem of volume $V$ will thus contain a number of spins
$N$ which is random and Poisson distributed:
$p(N) = \exp(-\rho V) (\rho V)^N / N! $.
This distribution must be used to weight the averaged free energy
so we are led to define
$ {\Xi^n} = \sum_N p(N){\overline Z_N^n}$.
By analogy with the statistical mechanics of pure systems, 
we shall call  this type of disorder ``grand canonical disorder''. 

The resulting theory is simpler than (\ref{eq:can}) and is defined by
\begin{eqnarray}
 {\Xi^n} &=& 
\exp(-\rho V)\int {\cal D} \phi_a 
[det J\beta]^{-{n\over 2}} \times\nonumber \\
&{\exp}& \Bigl( -{1\over 2 \beta}\int\!\int \sum_a 
\phi_a(r)J^{-1}(r-r')\phi_a(r')\ drdr' \nonumber \\ 
 &+& \rho\int \prod^n_a 2\cosh \phi_a(r) \ dr \Bigr).
\label{FieldT}\end{eqnarray}
Expanding the $\cosh$ one sees that the leading term 
corresponds to the random temperature or random mass, 
familiar from bond disordered approaches, and that depending
on the choice of interaction one might 
expect similar renormalisation group results\cite{DeLa1,DHSS}.

In order to relate this theory to measurable
quantities we return to
the original formulation of the model in equation (\ref{Hamiltonian})
and identify physical operators.
The spin density operator,
$M_a(r)=\sum \delta(r-r_i)S_i^a$ is closely related to the
field $\phi_a$ appearing in the theory.
The equations of motion following from the 
replicated version of (\ref{ZedN}) show that
the physical magnetisation density is given by,
\begin{equation} 
M  = {1\over V}\int \!dr \langle M_a(r)\rangle 
 = {1\over V}\langle\sum_i^N S_i^a \rangle\ 
={\langle\phi\rangle\over\beta\tilde J(0)},
\label{Meqn}\end{equation}
and the correlator 
$\langle M_a(r)M_b(r')\rangle $, is in terms of  
$G_{ab}(r-r') = \langle \phi_a(r)\phi_b(r') \rangle_{c}$, obtained as
\begin{equation} 
 \langle \tilde M_a(k) \tilde M_b(-k)\rangle_c
= {\tilde G^{ab}(k)\over\beta^2\tilde J^2(k)}
-{\delta^{ab}\over \beta \tilde J(k)} \ .
\label{Mcorr}\end{equation}
$M_a(r)$ is not, however, the operator sensitive to spin glass 
ordering, and it is natural to consider another operator 
$q_{ab}(r) = \sum_i \delta (r -r_a) S_i^a S_i^b$,
related to the non-linear susceptibility.
This new operator is composite and does not manifestly appear
in the field theory (\ref{FieldT}), it is for this reason that 
we must go beyond mean field theory to obtain non-trivial results.
Operators involving more spins 
can be introduced in the same way.


\bigskip

In the remainder of the letter
we analyse this field theory with the Gaussian variational method
which can be regarded as a generalisation of mean field theory.
This method, otherwise known as Hartree Fock, is a truncation of
the Schwinger Dyson equations and becomes exact in 
the limit of many spin components (such an m-component
theory is treated in a 
separate publication \cite{largeN}).
In the context of disordered systems, this method has has success in
calculating exponents for random manifolds \cite{GandM},
but one should bear in mind that important effects 
may occur at higher orders in $1/m$. 
In fact, for certain choices of the potential in the field theory
considered here, one can 
rigorously demonstrate a failure of the method \cite{DeLa1}.
We return to a discussion of the reliability of the approximation
at the end of the letter.

We allow the possibility of ferromagnetic order and make
the ansatze that 
$\langle \phi_a(r) \rangle = 
\bar\phi_a$ and $\langle \phi_a(r)\phi_b(r') \rangle_{c} = 
G_{ab}(r-r')$ (by translational invariance). 
The  variational free energy is given, 
up to constant terms, by
\begin{eqnarray}
n\beta F_{var}&=&
-{1\over 2} {\rm Tr} \log G_{ab}
+{1\over 2 \beta}{\rm Tr}\ G_{ab}J^{-1}  \\
&+&
{1\over 2 \beta}\sum_a \bar\phi_a^2\tilde J^{-1}(0) 
-{\rho}\Omega,\nonumber
\end{eqnarray}
where the above traces are both functional and on 
replica indices and where $\Omega$ is defined by
\begin{equation}
\sum_{S_a} {\rm exp}\left(\sum_a \bar\phi_aS_a
+{1\over 2}\sum_{ab} G_{ab}(0)S_aS_b \right)\ .
\label{omega}\end{equation}
As usual we do not expect 
breaking of replica symmetry on single-index objects and 
hence set $\bar \phi_a = \bar \phi$.
The variational equations are
\begin{equation} \bar\phi_a = \bar \phi = 
\beta\rho \tilde J(0)\Omega_a \end{equation}
and 
\begin{equation} \tilde G_{ab}^{-1}(k) = 
{1\over \beta}\delta_{ab}\tilde J^{-1}(k) 
- \rho \Omega_{ab}.\end{equation}
where $\Omega_a$ and $\Omega_{ab}$ are traces of the type (\ref{omega}) 
containing respectively $S_a$ and $S_a S_b$.


Within this approximation,
by introducing a source for the operator $q_{ab}(r)$ 
and using the FDT theorem, we obtain an equation 
for the correlation function 
$Q_{abcd} = \langle q_{ab}(r) q_{cd}(0) \rangle$. 
\begin{eqnarray}
\tilde Q_{abcd}(k)&=& \rho \Omega_{abcd}
+{\rho\over 2}\sum_{gh} \tilde \Sigma_{abgh}(k) 
\tilde Q_{ghcd}(k)\nonumber\\
\tilde\Sigma_{abgh}(k)&=& \sum_{ef}\Omega_{abef}\int\!{d^dp\over (2\pi)^d}
\tilde G_{eg}(p)\tilde G_{fh}(k-p)
\label{eq:fourindex},
\end{eqnarray}
where $\Omega_{abcd}$ is another object of the type (\ref{omega})
containing four $S$'s.


A replica symmetric (RS) ansatz for $G$ 
leads to a regime specified by two order parameters: 
the magnetisation $M$ (\ref{Meqn}) and the  Edwards Anderson 
order parameter $q_{EA}$. These parameters are determined
by a pair of equations very similar to the
mean field equations for the Sherrington Kirkpatrick (SK) model\cite{SK}
\begin{eqnarray}
M
&=&{\rho\over \sqrt{2\pi}}\int\! d\xi e^{-{\xi^2\over 2}}
\tanh\left(\beta\tilde J(0)M +  \xi\sqrt{g_1})\right)\nonumber\\
q
&=&{1\over \sqrt{2\pi}}\int\! d\xi e^{-{\xi^2\over 2}}
{\rm tanh}^2\left(\beta\tilde J(0)M + \xi\sqrt{g_1})\right)
\label{RSeqns}
\end{eqnarray}
Where $g_1$ (the off diagonal part of $G_{ab}(0)$) is given by
\begin{equation}
g_1
= \rho \beta^2 q\int\! {d^dk\over (2\pi)^d}
{\tilde J^2(k)\over \left(1-(1-q)\rho\beta \tilde J(k)\right)^2}
\end{equation}
The simplest solution of these equations yields
the high temperature, low density 
paramagnetic region with $q_{EA}=0$ and $M = 0$. 
In this region, 
the two index correlator (\ref{Mcorr}) is known and 
equation (\ref{eq:fourindex}) can be solved for
the most interesting correlator,
$\langle q_{ab}(k) q_{ab}(-k) \rangle$ 
\begin{equation}
{\rho\beta^2\over 1-\rho \int\! {d^dp\over (2\pi)^d}
{\tilde J(k) \tilde J(p-k)\over 
\left(1-(1-q)\rho\beta \tilde J(k)\right)
\left(1-(1-q)\rho\beta \tilde J(p-k)\right)}}
\label{eq:qcorr}\end{equation}
This correlator is simply related to
$\overline{ \langle S(k) S(-k)\rangle^2}
= \sum^\prime_{ab}\langle q_{ab}(k) q_{ab}(-k) \rangle$ 
and the divergence in the above formula
signals the onset of a spin glass phase.
The divergence occurs on a line in the temperature density
plane specified by $g_1 = q$ and 
coincides
with the AT line as determined by stability considerations \cite{DeLa1,AlTh}.
Furthermore the phase boundary also coincides with the line on which
the RS equations (\ref{RSeqns}) develop
solutions with non-zero $q$.
This situation also occurs in the
SK model and suggests a continuous
breaking of replica symmetry.


We shall look for continuous replica symmetry broken solutions and
parameterise the off-diagonal part of the matrix $\tilde G_{ab}(k)$
by a continuous Parisi function $\tilde g(k,u)$
where $u\in [0,1]$, and a  diagonal part  denoted by $\tilde g_D(k)$.
For such a matrix, $\Omega$ (\ref{omega}) 
is very similar to the free energy in the SK model; it cannot 
be obtained in a closed form and a standard strategy is to
work close to the transition line by expanding $\Omega$ up to a term of 
$O(g^4)$ which in the SK model is the first term leading to a breaking
of replica symmetry. The expansion is \cite{PytteR}
\begin{equation} 
{\Omega-1\over n} \approx  {1\over 2} g_D(0) 
- {1\over 4} \int_0^1 \left( 
g^2 + {1\over 6} g^4 -{u\over 3} g^3 - g\int_0^u g^2 \right) \ du
\end{equation}
The remaining terms in the action are easily computed within the algebra
of Parisi matrices \cite{GandM}. The variational equations one obtains are
\begin{eqnarray}
[\tilde g_D(k)]^{-1} &=& + \rho \sigma_D(k) 
= \rho \left( {\tilde J^{-1}\over \beta\rho} - 1 \right) \cr
[\tilde g(u,k)]^{-1} &=& - \rho \sigma (u) 
= 2\rho {\delta \Omega\over \delta\tilde g}
\label{eq:rsb1}\end{eqnarray}
Defining
\begin{eqnarray}
D_D(k) &=& \sigma_D + \langle \sigma \rangle \cr 
D(u,k) &=& \sigma_D + \langle \sigma \rangle + [\sigma ](u),
\end{eqnarray}
(in the notation of \cite{GandM}), the equations can
be inverted to find
\begin{eqnarray}
\tilde g_D(k) &= {1\over \rho D_D(k)} \left( 1 + \int_0^1\! {du\over u^2}
 {[\sigma ](u) \over D(u,k)} + {\sigma(0)\over  D_D(k)}\right)\cr
\tilde g(u,k) &= {1\over \rho D_D(k)} 
\left( {[\sigma ](u) \over u D(u,k)} + \int_0^u\!{dv\over v^2}
 {[\sigma ](v) \over D(v,k)} + {\sigma(0)\over  D_D(k)}\right).
\end{eqnarray}
Proceeding by differentiating 
the second equation of (\ref{eq:rsb1}) with respect to 
$u$ one obtains $\sigma ' = 0$ or
\begin{equation}
\left( 1+  g^2 - u g  - \int_u^1 g \right) = 
\rho \left( \int\!  {d^dk\over (2\pi)^d} {1\over D^2} \right)^{-1}
\label{eq:rsb2}
\end{equation}
Taking a second derivative in some region where equation
(\ref{eq:rsb2})
holds we find:
\begin{equation}
g = \alpha (u) u 
= {u\over 2} \left( 1 + 2\rho^2 
{\int  {d^dk\over (2\pi)^d} {1\over D^3} \over 
\left( \int  {d^dk\over (2\pi)^d} {1\over D^2} \right)^3} \right).
\end{equation}
For 4 or more dimensions, in the limit
in which the short distance cutoff is removed,
this equation is simple and we find a
scenario similar to that found in the SK model.
In general the function $\alpha(u)$ depends on $\sigma(u)$ and one obtains a 
first order nonlinear differential equation for $g(0,u)$. 
In all cases we have considered, the power series solution near the
origin starts with a linear term. 
For consistency with our perturbative analysis the region where 
$g(0,u)$ is non constant must be close to the origin. 
More precisely there must be a break point $u_0$ with small value,
above which $g(0,u)$ is constant and equal to $\alpha u_0$ if the solution 
is to be continuous. The breaking pattern scenario 
is reminiscent of the random manifold problem
with long range disorder \cite{GandM} and is
qualitatively the same as  found in the SK model near $T_c$\cite{Pa}.


It is useful to illustrate these results for
a specific interaction and
for the purposes of this letter
we consider an RKKY-like oscillatory potential in 3-dimensions.
\begin{equation} \tilde J(k) = 
\mu^{-3}\theta(\mu - |k|) .\end{equation}
The dimensional constant $\mu$ merely sets the scale of the problem and
can be set to 1. Using equation (\ref{eq:qcorr})
we obtain for $\langle q_{ab}(k) q_{ab}(-k) \rangle$ 
in the paramagnetic phase
\begin{equation}
\tilde Q_{abab}=\cases
{{\rho\over 1 - {\rho\over 96\pi^2} \left({\beta\over1-\rho\beta}\right)^2
(k+4)(k-2)^2},
 & for \hbox{$k < 2$};\cr
\rho, & for \hbox{$ k> 2$}.\cr}
\end{equation}
The spin glass phase boundary is given by 
$\rho = (1 + 12\pi^2 T-\sqrt{1+24\pi^2 T})/(12\pi^2)$,
and the exponent associated with the transition is $\eta =1$.
Numerical inspection of the RS equations finds
stable ferromagnetic solutions at low temperature
because at very high densities the positive short range part of
the potential can dominate.
We illustrate the expected form of the phase diagram in figure 1.

The expansion just below the AT line gives rise to a differential
equation as described above, the leading solution at small $u$
being linear. The break point $u_0$ can be calculated in terms of
the deviation from the AT line:
$u_0 = \delta\beta {2\over 3\pi^2} 
{\rho\beta\over(1+\rho\beta)(1-\rho\beta)}$.
Despite having the full  structure of the two
index correlators $\tilde g(k,u)$, leading to non trivial 
momentum dependence in $\tilde g_D(k)$ related to the connected
magnetic correlation function (\ref{Mcorr}), the analysis only holds
close to the spin glass transition 
and is unable to address the ferromagnetic transition 
which takes place at much lower temperature.
The four index correlation functions $Q_{abcd}$
contain much of the physics of the spin glass phase:
for example the $\theta$-exponent \cite{FiHu}
may be extracted from the long distance behaviour of such objects. 
Equation (\ref{eq:fourindex}) is however an equation carrying 
four replica indices and the solution in the case of continuous 
replica symmetry breaking is technically rather formidable
requiring extensions of the methods described in \cite{Kondor}.


\bigskip

As we have emphasised, we have used the Gaussian variational method 
because it is the simplest generalisation of mean field theory that
gives us access to spin glass physics. We now discuss the
reliability of the approximation. 
Certainly we should expect it to be exact for $m$-component
Heisenberg spins in the limit $m\to \infty$.
In this case \cite{largeN} 
we obtain a similar picture to that described above:
namely a high temperature phase separated from a spin glass
phase at low temperature. The form of the spin glass phase
is  RS with $q$ non-zero and the equations for $Q_{abcd}$ 
may be solved to find that $Q_{abab}$ stays
critical below the transition with exponent $\theta$
given by $\theta = d-1$ for RKKY-like interactions \cite{foot}.
Applying the method to $m=1$ Ising spins will lead to
errors, but we hope that certain features will be correct.
A case that can be analysed rigorously is that of
a purely ferromagnetic interaction and Ising spins
where we can demonstrate that the spin glass and ferromagnetic transitions 
must be simultaneous \cite{DeLa1}. The Gaussian variational method fails
in this respect, predicting a spin glass transition at slightly 
higher temperature than the ferromagnetic transition.
Indeed this effect has been noticed before, and Sherrington \cite{Sh1} 
has identified relevant diagrams that are ignored in the 
GV approach. 

Another shortcoming, not related to the Gaussian variational
approximation, may
also be present in our treatment of spin glass ordering. That is
that we have only taken our analysis as far as order parameters
with two replica indices which is known, for example in the 
Viana Bray model, not to be correct \cite{ViBr}. 
This effect may be apparent in the case of an antiferromagnetic
interaction. There is no difficulty of principle in extending our methods
to consider operators with more replica indices, but in practice, the 
calculations soon become unwieldly.


We would like to acknowledge useful discussions with J.P.~Bouchaud,
M.~Ferrero, G.~Iori, J.~Ruiz-Lorenzo, M.~M\'ezard,  R.~Monasson, 
T.~Nieuwenhuizen and G.~Parisi.



\begin{figure}
\epsfxsize=230pt\epsffile{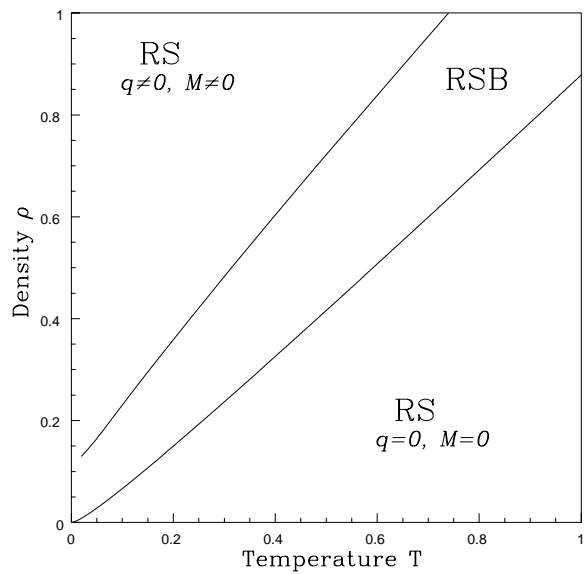}
\caption{
Phase diagram for the 3 dimensional RKKY interaction
using Hartree approximation.
}
\protect\label{figone}
\end{figure}


\begin{references}
\bibitem{foot}{The bounds given in \cite{FiHu} are for Ising spins.}
\end{references}
\end{document}